\documentclass[aip, preprint]{revtex4-1}
\usepackage{graphicx}
\usepackage{xcolor}

\begin{document}

%\preprint{}

\title{First radiative shock experiments on the SG-II laser} %Title of paper



\author{Francisco Suzuki-Vidal}
\email[]{f.suzuki@imperial.ac.uk}
%\thanks{}
\affiliation{Blackett Laboratory, Imperial College London, London SW7 2BW, UK}

\author{Thomas Clayson}
\affiliation{First Light Fusion Ltd., Yarnton OX5 1QU, UK}
\altaffiliation{Current affiliation: Magdrive Ltd, Harwell OX11 ORL, UK}

\author{Chantal Stehl\'{e}}
\affiliation{LERMA, Sorbonne-Universit\'{e}, Observatoire de Paris, CNRS, France}

\author{Uddhab Chaulagain}
\affiliation{ELI Beamlines Center, Institute of Physics, Czech Academy of Sciences, Za Radnici 835, Dolni Brezany, Czech Republic}

\author{Jack W.D. Halliday}
\affiliation{Blackett Laboratory, Imperial College London, London SW7 2BW, UK}

\author{Mingying Sun}
\affiliation{Shanghai Institute of Optics and Fine Mechanics, Chinese Academy of Sciences, China}

\author{Lei Ren}
\affiliation{Shanghai Institute of Optics and Fine Mechanics, Chinese Academy of Sciences, China}

\author{Ning Kang}
\affiliation{Shanghai Institute of Optics and Fine Mechanics, Chinese Academy of Sciences, China}

\author{Huiya Liu}
\affiliation{Shanghai Institute of Optics and Fine Mechanics, Chinese Academy of Sciences, China}

\author{Baoqiang Zhu}
\affiliation{Shanghai Institute of Optics and Fine Mechanics, Chinese Academy of Sciences, China}

\author{Jianqiang Zhu}
\affiliation{Shanghai Institute of Optics and Fine Mechanics, Chinese Academy of Sciences, China}

\author{Carolina de Almeida Rossi}
\affiliation{Blackett Laboratory, Imperial College London, London SW7 2BW, UK}

\author{Teodora Mihailescu}
\affiliation{Blackett Laboratory, Imperial College London, London SW7 2BW, UK}

\author{Pedro Velarde}
\affiliation{Instituto de Fusi\'{o}n Nuclear Guillermo Velarde, Universidad Polit\'{e}cnica de Madrid, 28006 Madrid, Spain}

\author{Manuel Cotelo}
\affiliation{Instituto de Fusi\'{o}n Nuclear Guillermo Velarde, Universidad Polit\'{e}cnica de Madrid, 28006 Madrid, Spain}

\author{John M. Foster}
\affiliation{AWE plc., Aldermaston, Reading, Berkshire RG7 4PR, UK}

\author{Colin N. Danson}
\affiliation{AWE plc., Aldermaston, Reading, Berkshire RG7 4PR, UK}

\author{Christopher Spindloe}
\affiliation{Science and Technology Facilities Council, Rutherford Appleton Laboratory, Harwell Campus, Chilton, Didcot, Oxon OX11 0QX, UK}

\author{Jeremy P. Chittenden}
\affiliation{Blackett Laboratory, Imperial College London, London SW7 2BW, UK}

\author{Carolyn Kuranz}
\affiliation{University of Michigan, Ann Arbor, MI 48109, USA}


\date{\today}

\begin{abstract}

We report on the design and first results from experiments looking at the formation of radiative shocks on the Shenguang-II (SG-II) laser at the Shanghai Institute of Optics and Fine Mechanics in China. Laser-heating of a two-layer CH/CH-Br foil drives a $\sim$40 km/s shock inside a gas-cell filled with argon at an initial pressure of 1 bar. The use of gas-cell targets with large (several mm) lateral and axial extent allows the shock to propagate freely without any wall interactions, and permits a large field of view to image single and colliding counter-propagating shocks with time resolved, point-projection X-ray backlighting ($\sim20$ $\mu$m source size, 4.3 keV photon energy). Single shocks were imaged up to 100 ns after the onset of the laser drive allowing to probe the growth of spatial non-uniformities in the shock apex. These results are compared with experiments looking at counter-propagating shocks, showing a symmetric drive which leads to a collision and stagnation from $\sim$40 ns onward. We present a preliminary comparison with numerical simulations with the radiation hydrodynamics code ARWEN, which provides expected plasma parameters for the design of future experiments in this facility.

\end{abstract}


\maketitle %\maketitle must follow title, authors, abstract and \pacs


\section{Introduction}
\label{Introduction}

Radiative shocks are formed when radiative losses from the shock can modify its structure. This occurs when the radiative energy flux is comparable to the kinetic energy flux at the shock front. In this regime, radiation can modify both the pre- and post-shock regions. Radiative effects increase with the shock speed due to stronger post-shock heating and, in a first approximation for typical experimental conditions, radiative effects start playing a role at shock velocities of 10$-$100's km/s and gas pressures of $\sim$0.1$-$1 bar [\cite{Drake2005a}]. With present day laser facilities it is possible to reach such shock speeds by compressing and focusing the laser energy into a solid (e.g. a piston) or gas target. This has been done in a number of different laser facilities with a variety of energies, intensities (typically $\gtrsim$ 10$^{14}$ W/cm$^2$) and configurations, e.g.~point-explosions in a gas background, shock tubes and gas-cells (a recent review can be found in [\cite{Drake2019}]). 

Recent works have looked at bridging the gap between experiments and theory/numerical simulations of radiative shocks [\cite{Mabey2020}] and applications to astrophysics [\cite{Mabey2020apj}]. In particular, recent experiments have looked at the interaction of a piston driven shock with an obstacle [\cite{Michel2018HPLSE}, \cite{Michel2019apj}]. However, several issues have lead to difficulties to make a complete bridge between simulation and experiments, for instance the question of opacity for heavy gases (e.g. xenon) or the nature of the rise of instabilities and the role played by radiation. In addition, at higher velocity, temperature increases strongly and non-LTE effects start to play a role [\cite{Rodriguez2015PRE}]. It is, therefore, of key importance to continue experimental efforts to obtain more experimental data to be compared with theoretical works.

The experiments presented here use the Shenguang-II (SG-II) laser to drive shocks via piston-action from a foil attached to one of the ends of a gas-cell target. Although the SG-II laser has been operational for many years, future improvements planned for this facility in the coming years make these first experiments critical for planning and testing of future experimental campaigns. 

The targets are characterised by a large internal volume and field of view to probe the dynamics of the shock as a function of time without any shock-wall interactions. This configuration is similar to first experiments performed on the Orion laser, where shocks were driven in xenon [\cite{Suzuki-Vidal2017PRL}] and neon [\cite{Clayson2017HEDP}] with similar laser energies and gas pressures. However, the SG-II targets have a larger diagnostic field compared to the Orion experiments and, through the use of argon, we aim to study the formation of spatial non-uniformities in the shocks which were only investigated preliminarily at Orion.

% FIG1: EXPERIMENTAL SETUP WITH LASER DRIVE AND DIAGNOSTICS
\begin{figure*}[t!]
\centering
\includegraphics[width = 16cm]{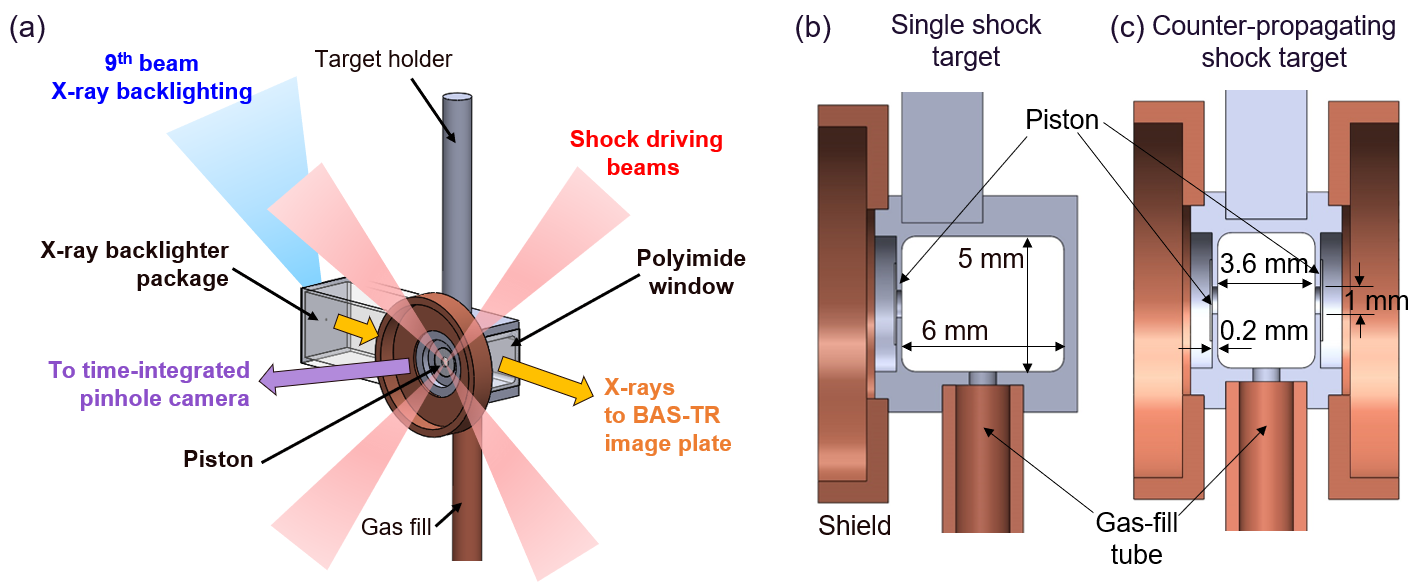}
\caption{(a) Schematic diagram of the experimental setup for a single shock target on the SG-II laser, with a similar configuration used for counter-propagating shocks. (b)-(c) Cross-section of single and counter-propagating shock targets.}
\label{FIG01_target_setup}
\end{figure*}

\section{Experimental setup}

The SG-II laser [\cite{Zhu2018HPLSE}] can drive up to eight beams with a pulse duration of 1 ns in two opposite directed groups of four beams. Thus, four to eight beams were used depending on whether the experiment was aimed at producing single or counter-propagating shocks. The overall experimental setup is shown schematically in Fig.~\ref{FIG01_target_setup}a for a single shock target. The targets are similar for the case of counter-propagating shocks, where 4 additional beams are focused on a second piston placed opposite to the first one. Detailed dimensions for both types of targets are shown in Figs.~\ref{FIG01_target_setup}b-c.

\begin{figure}
\centering
\includegraphics[width = 10cm]{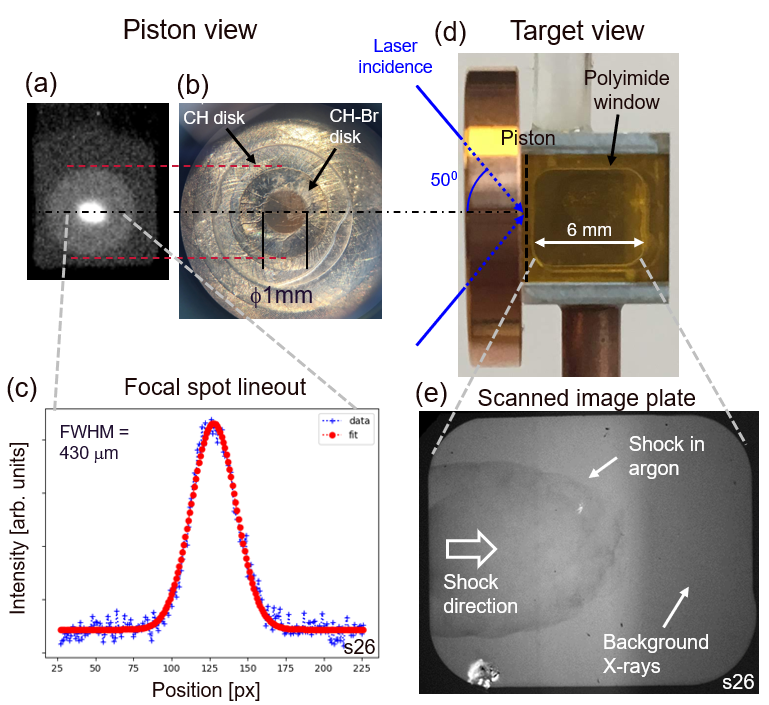}
\caption{(a)-(c) View of the piston: (a) X-ray emission from the laser-piston interaction from a time-integrated pinhole camera diagnostic. (b) Microscope image of the 1 mm diameter shock aperture and CH-Br / CH pistons. (c) Lineout of (a) and Gaussian fit to estimate the laser spot size. (d)-(e) Target view: Field of view  of the X-ray backlighting diagnostic of a single shock target and example of raw X-ray image result respectively.}
\label{FIG02_piston}
\end{figure}

The lasers driving each shock had a total energy of $\sim$1 kJ, a top-hat temporal profile with a duration of $\sim$1 ns and a frequency of 3$\omega$ ($\lambda$ = 351 nm). The laser beams were focused to a nominal spot diameter of $\sim$300 $\mu$m with a super-Gaussian spatial profile with $n\sim4-6$. Measured laser parameters for the experiments are presented in the Results section. An external laser beam (the $9^{th}$ beam) was used to drive X-ray backlighting onto a scandium foil to image the shock inside the gas-cell side-on, i.e.~normal to the direction of propagation of the shock. The backlighter beam had an energy $E_{9th}\sim500$ J and a pulse duration of $t_{9th}\sim1$ ns, with a spot size of $\phi_{9th}\sim150$ $\mu$m at $3\omega$. In addition to the X-ray backlighter diagnostic, a filtered, time-integrated pinhole camera recorded the X-ray emission from the laser-piston interaction onto an image plate to estimate the size of the focal spot during each shot. An example of results from the two diagnostics is presented in Fig.~\ref{FIG02_piston}.

\subsection{Target design}

Gas-cell targets were designed for SG-II with two main purposes: 1) To provide a large ambient gas volume in order to allow the shocks to propagate without being subject to radiative or hydrodynamic interactions with the gas cell-walls, 2) To maximize the X-ray backlighter diagnostic field of view in order to follow the evolution of the shocks for times up to 100 ns.  

Two target designs (Fig.~\ref{FIG01_target_setup}b) were fielded on SG-II: A single-shock target with a diagnostic window size (width$\times$height) of 6$\times$5 mm$^2$, and a counter-propagating shock target with a diagnostic window size of 3.6$\times$5 mm$^2$. The window width of the counter-propagating shock targets was constrained by the maximum separation between the two opposite groups of long-pulse laser beams, i.e.~4 mm between focal spots. The gas-cell windows were sealed with a 25 $\mu$m thick polyimide foil, with a transverse distance between windows of 8 mm. The targets were positioned inside the chamber with a 3 mm diameter plastic rod glued at the top of each target, connected to a 5-axis target positioner. 

The targets for SG-II were designed based on previous experiments on the Orion laser [\cite{Spindloe2017HPLSE}], with several improvements implemented for SG-II: 1) a thinner, 0.2 mm thick frame between the piston and the edge of the diagnostic window to probe the early-time behaviour of the shock (cf. 1 mm for Orion). A drawback of decreasing the frame thickness was that this reduced the shielding which the target provided to prevent hard X-ray emission from the laser-piston interaction from contaminating the signal on the X-ray backlighter diagnostic. This meant the level of background noise in diagnostic images from these experiments was higher than the level observed previously on Orion. 2) A smaller, 1 mm diameter aperture for the piston was in contact with the gas (cf.~3$-$5 mm in Orion) to reduce the `swelling' of the piston with gas-fill pressure inside the vacuum chamber. This aids to achieve a more consistent laser focal spot size during each shot. 3) Lastly, the SG-II targets were gas-filled in situ while inside the vacuum chamber, allowing for a faster shot turnaround and accurate gas-pressure measurement right before each shot. This gas-fill system was also used in previous experiments on the PALS laser [\cite{Chaulagain2015HEDP}, \cite{Singh2017HEDP}]. For the experiments presented here, argon with an initial gas pressure of $P_{Ar}$=1 bar was used ($\rho_{Ar}$=1.67 mg/cc). 

The pistons for SG-II were made with a layer of 30 $\mu$m thick CH (on the laser-drive side, $\phi\sim3.2$ mm diameter, nominal mass density $\rho_{CH}$=0.9 g/cc) followed by a 50 $\mu$m thick CH-Br layer (on the gas side, diameter $\phi\sim2$ mm, nominal mass density $\rho_{CH-Br}$=1.53 g/cc). The brominated plastic layer helped shielding the gas from X-rays produced in the laser-CH interaction in order to reduce radiative pre-heating.

% FIG02: TARGET PISTON + TARGET XRBL EXAMPLE

\subsection{X-ray backlighter diagnostic}

A point-projection X-ray backlighter (XRBL) package was attached to the side of each target. It consisted of a 5 $\mu$m thick, 500 $\mu$m diameter scandium foil (a ‘microdot’) which was supported on a 100 $\mu$m thick tantalum substrate, glued to a 3-D printed acrylic frame that was aligned to the center of the gas-cell window. The X-ray source size was constrained by a $\phi_{PH}=$20 $\mu$m diameter pinhole which was laser-cut into the Ta substrate. This pinhole was coated with a 9 $\mu$m Parylene-N layer which served to prevent the closure of the pinhole by the plasma which was formed in the interaction between the backlighter beam and the Sc foil [\cite{Kuranz2006}]. Laser etched markings were added to the Ta substrate and the Sc foil to ensure accurate alignment during target manufacture and target alignment before each shot.

The backligher beam average laser parameters were an energy of $E_{9th}=478\pm71$ J and pulse duration $t_{9th}=1042\pm188$ ps. The nominal spot size was $\sim$150 $\mu$m, giving an intensity of $I_{9th}\sim 2.6\times10^{15}$ Wcm$^{-2}$. Under these drive conditions, the X-ray emission from the interaction of the backlighter beam with the Sc foil is dominated by He-alpha emission with a photon energy of 4.3 keV [\cite{Ruggles2003RevSci}], providing a quasi-monoenergetic source of X-rays to study the plasma in the shocks. The X-rays were recorded in a film-pack with Fuji BAS-TR image plate, filtered with a 12.5 $\mu$m thick titanium filter and 2$-$3 layers of 8 $\mu$m thick aluminised polypropylene for optical and debris shielding. The XRBL package was placed at a distance of $p=21$ mm from the axis of the gas-cell, with the image plate placed at a typical distance of $q\approx226$ mm, thus resulting in a point-projection magnification of $M=(p+q)/p\approx$12.

The spatial resolution of the XRBL was estimated by illuminating a vanadium grid attached to a target and fitting the resulting spatial profile to the convolution of the ideal point projection from the grid with a Gaussian function. This results in a resolution of $\sim19$ $\mu$m, which is compatible with the size of the pinhole ($\phi_{PH}=20$ $\mu$m) and the geometrical resolution given by $\delta\approx\phi_{PH}(M-1)/M\sim$18 $\mu$m [\cite{Sinars2004IEEE}]. The field of view of the XRBL diagnostic for a single shock target together with an example of results is shown in Fig.~\ref{FIG02_piston}d-e. It should be noted that for the shock speed of $\sim$40 km/s, the motion blurring for a 1 ns exposure XRBL is $\sim$40 $\mu$m, about twice the size of the pinhole.

\section{Results}

\begin{figure*}[t!]
\centering
\includegraphics[width = 16cm]{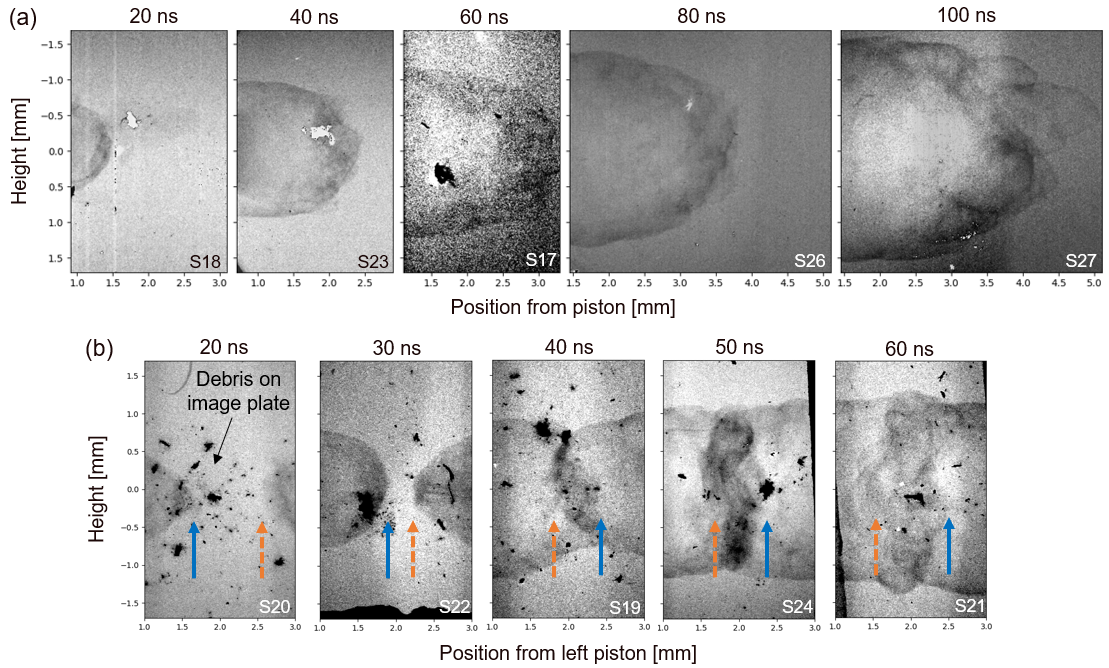}
\caption{X-ray backlighting results for (a) single and (b) colliding shocks. Artefacts from hard X-ray background (see Fig.~\ref{FIG02_piston}e) have been removed for visual purposes. For colliding shocks, the position is taken relative to the left-hand side piston and the vertical arrows mark the approximate position of the shock fronts in each frame. The arrows suggest that the shocks interpenetrate however, in reality, the experiments are in a collisional regime where no interpenetration occurs.}
\label{FIG03_results}
\end{figure*}

Results from single and colliding radiative shock experiments are presented in Figs.~\ref{FIG03_results}a-b respectively. Each image was obtained from a different shot with similar initial conditions for the laser drive and gas-fill pressure. Overall, the results showed good shot-to-shot reproducibility for the shock dynamics. Average laser-drive parameters in these experiments were an energy of $E_{laser}=960\pm53$ J and a pulse duration of $t_{laser}=1039\pm63$ ps. The spot size was estimated from the full-width half-maximum of Gaussian fits from X-ray self-emission recorded with the time-integrated pinhole camera (Fig.~\ref{FIG02_piston}c), resulting in an oval spot with size $\phi_{laser}\approx424\pm11\times324\pm15$ $\mu$m (width$\times$height), thus an average laser intensity of $I_{laser}\sim8.6\times10^{14}$ Wcm$^{-2}$. The laser intensity could be lower than this measured value due to the spot size images being time-integrated.

Dark spots in the images are due to debris reaching the image plate, which appear to be more pronounced in the case of colliding shocks. The XRBL results were characterised by vertical bands with abrupt changes in X-ray intensity which are attributed to background X-ray emission coming from the laser-piston interaction (see the right side of Fig.~\ref{FIG02_piston}e). These artefacts were removed from the images in Fig.~\ref{FIG03_results} by subtracting the average intensity distribution from regions where the shock was not present.  Future experiments will look at reducing debris and background emission on the image plate by adding extra shielding on the targets. 

Fig.~\ref{FIG03_results}a shows the results for a single shock for times between 20$-$100 ns, with each image corresponding to a separate experiment with nominally identical initial conditions. The shocks show a semi-hemispherical shape with a good degree of axial symmetry, i.e.~respect to a height of 0 mm in the window. In the earliest image, at 20 ns, the shock front is seen as a fairly smooth feature, however from 40 ns onwards the shock develops spatial non-uniformities which grow in size as time increases. Rough wavelengths of these features are $\lambda\sim250$ $\mu$m at 80 ns and $\lambda\sim500$ $\mu$m at 100 ns. 

Fig.~\ref{FIG03_results}b shows the results for two colliding shocks in a counter-propagating configuration. Before and after the collision, the two shocks show a good degree of left-right symmetry and thus demonstrate a well controlled laser drive and a consistent target fabrication procedure. These results show similar dynamics to previous experiments on the Orion laser using xenon at the same initial gas mass density ($\rho_{Xe}\sim$1.6 mg/cc) [\cite{Suzuki-Vidal2017PRL}], thus proving that is possible to carry on and improve similar experiments of this type in SG-II in the future.

% FIG03: RESULTS XRBL FIGURE WEEK 1 (single and double shocks separately)

% FIG04: Vz from week 1 - single vs double
\begin{figure}
\centering
\includegraphics[width = 8.5cm]{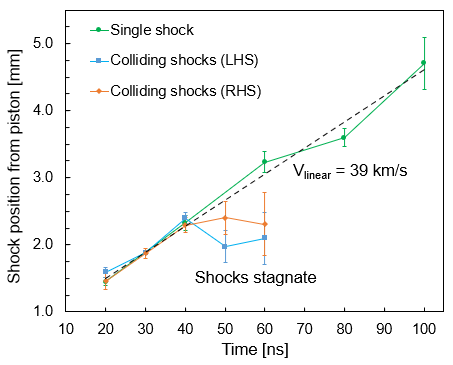}
\caption{Shock front position as a function of time measured from the results in Fig.~\ref{FIG03_results} for single and colliding shocks. For the latter, the positions of the shock fronts are marked in Fig.~\ref{FIG03_results}b with matching colours.}
\label{FIG04_Vz_week1}
\end{figure}

Fig.~\ref{FIG04_Vz_week1} shows the position of the tip of the shock fronts with respect to the initial position of the piston for single and colliding shocks, measured from the data shown in Fig.~\ref{FIG03_results}. For visual purposes, the position of the shocks is marked at 40, 50 and 60 ns as if the shocks interpenetrated, however the argon plasma in the shocks is expected to be in the collisional regime as the ion mean free path of argon is $\lambda_{mfp}\lesssim1$ $\mu$m [\cite{Park2012HEDP}]. This is estimated from plasma parameters from initial simulations presented later in the paper together with an estimate of the average ionization of argon for these parameters from [\cite{Espinosa2017PRE}]. Fig.~\ref{FIG04_Vz_week1} shows that the shock front position is fairly indistinguishable between single and colliding shocks up until 40 ns, i.e.~right at the time of the collision. The collision leads to the stagnation of the shocks which is supported by the ram pressure of the piston material behind the shock front. For a single shock, a linear fit of the shock position as a function of time results in a velocity of $V_{shock~front}\sim39$ km/s.

\begin{figure*}
\centering
\includegraphics[width = 16cm]{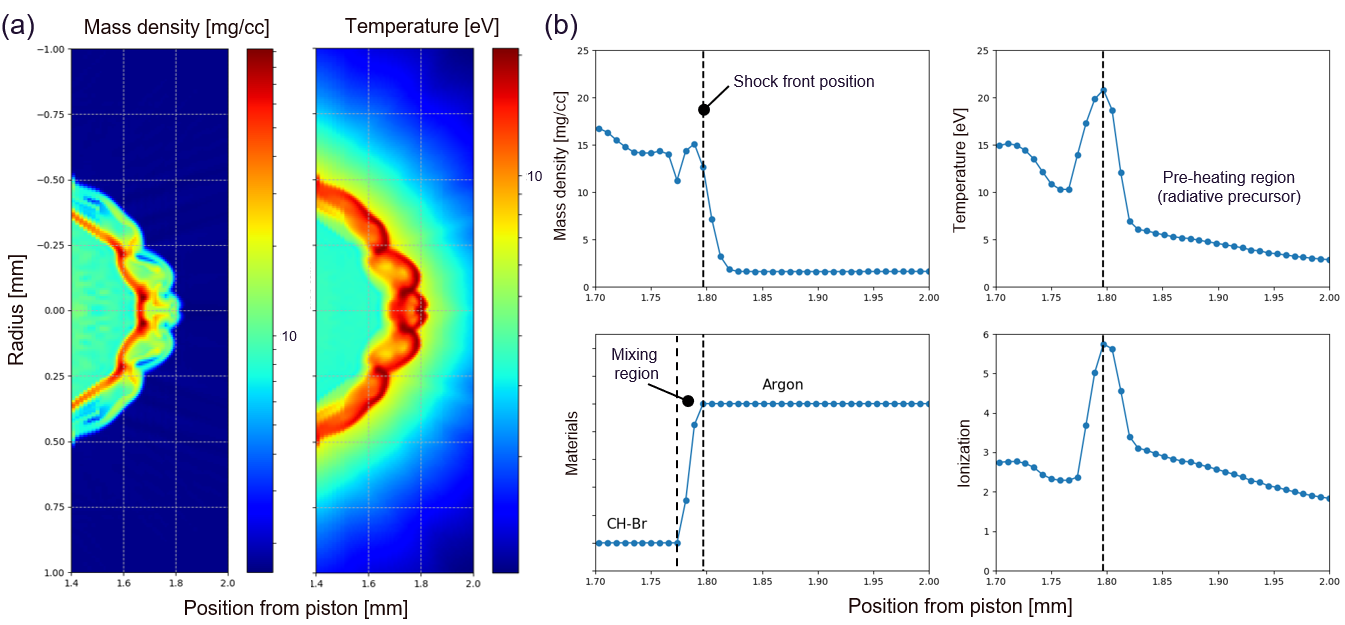}
\caption{Numerical simulations of the experiments with the 2-D radiation hydrodynamics code ARWEN at 20 ns. (a) Maps of mass density and temperature. (b) Axial lineouts (at a radius of 0 mm) of mass density, temperature, materials and ionization from (a).}
\label{FIG05_sims_week2}
\end{figure*}

To complement the experimental results, preliminary numerical simulations of the experiments are presented in Fig.~\ref{FIG05_sims_week2} for single shock at 20 ns. The simulations were performed with the 2-D radiative hydrodynamics code ARWEN [\cite{OgandoVelarde2001}, \cite{Garcia-Senz2019}]. A full comparison with the experimental data requires further testing of these simulations with several initial conditions, thus this single output is used to infer characteristic plasma conditions and make first estimates and further work will look at presenting a detailed simulation study. These first simulations were obtained using the initial conditions in the experiments, i.e.~an initial argon pressure of $P_0=1$ bar ($\rho_0=1.67$ mg/cc), a laser drive energy of $E_{laser}=962$ J, a duration of $t_{laser}=1$ ns and a focal spot of diameter $\phi_{laser}=370$ $\mu$m. The resolution of the simulations was 7 $\mu$m. 

Fig.~\ref{FIG05_sims_week2} shows 2-D maps of mass density and temperature together with axial (at a radius of r=0 mm) lineouts of these quantities together with materials and ionisation. The profiles of density, temperature and ionisation are representative of the typical structure of a radiative shock: a sharp jump in temperature and density at the shock front position (shown as a vertical dashed line at $\sim$1.79 mm), followed by a decrease in temperature and an increase in density post-shock, which we identify as the \textit{cooling region}. The post-shock region extends for $\sim$20 $\mu$m with a characteristic mass density of $\rho_{post-shock}\sim12-15$ mg/cc resulting in a density jump of $\rho_{post-shock}/\rho_0\sim12/1.67\sim7$, of the order of the density jump of 4 expected for a strong shock. The profile of materials post-shock shows a mixing region of CH-Br and Ar extending up to $\sim$1.77 mm. The temperature in the shock peaks at $T\sim20$ eV, and the temperature profile is characterised by a pre-heating region ahead of the shock with $T\sim2-5$ eV, which can be identified as the \textit{radiative precursor}. 

% FIG05: ARWEN sims week 2 - ONLY20ns

\section{Discussion and conclusions}

We have presented first results from experiments looking at the formation of radiative shocks in argon with piston-driven gas-cells on the SG-II laser. The main diagnostic fielded was point-projection X-ray backlighting which, combined with a new gas-cell design, allowed the study of shock evolution for single and counter-propagating colliding shocks. In the case of colliding shocks, the results are similar to previous experiments on the Orion laser, demonstrating the feasibility of this platform on SG-II to carry on future experiments of this type. The study of the post-shock region in single shocks had not been looked at in detail in previous experiments, e.g. on Orion and other similar laser facilities. For instance, work by [\cite{Michel2018HPLSE}] and references therein have mostly focused on the radiative precursor region ahead of the shock and rarely studied the dense post-shock region. 

One of the main results of these experiments on SG-II was to study the morphology of a single shock up to 100 ns to understand its evolution. The shock is characterised by the growth of spatial non-uniformities, with typical wavelenghts $\lambda\sim250-500$ $\mu$m. Our first numerical simulations with the code ARWEN show the formation of similar features at 20 ns with a rough wavelength of $\lambda\sim125$ $\mu$m in line with the experimental results, however it is not clear how sensitive are these features to different initial conditions in the simulations and how can these features change if the resolution of the XRBL diagnostic is improved.

We can estimate characteristic timescales for the growth of hydrodynamic instabilities from a shocked- clump model presented in [\cite{Hansen2017apj}]. For a strong shock, the Rayleigh-Taylor instability is expected to grow in a timescale given by $t_{RT}\sim\frac{\sqrt{\eta}~r_{spot}}{v_{shock}}$ with $\eta$ the ratio of post-shock to pre-shock density, $r_{spot}$ the radius of the laser focal spot and $v_{shock}$ the tip shock velocity. From the simulations at 20 ns we estimate $\eta\sim12/1.67\sim7$ and from the experiments $r_{spot}\sim185$ $\mu$m and $v_{shock}\sim39$ km/s, resulting in $t_{RT}\sim5$ ns in line with the experimental results. Similarly, the growth of the non-linear thin shell instability (NLTSI) is given by the fragmentation time $t_{frag}\sim \frac{h}{C_s}$ with $h$ the thickness of the shocked layer and $C_s$ its sound speed. For shocked argon at a temperature of $T\sim20$ eV, $C_s\sim9$ km/s, and for a shock thickness of $h\sim100$ $\mu$m this results in $t_{frag}\sim10$ ns, again consistent with the timescales observed in the experiments presented here.

The role of radiative cooling in the shock can be estimated using the plasma parameters from the simulations of $\rho\sim10$ mg/cc and $T\sim20$ eV. The expected cooling time for argon for these conditions (see Fig.~4 in [\cite{Espinosa2017PRE}]) is $t_{cool}\sim1$ ns, indicating that the plasma in the shock is radiatively cooled and thus could be prone to the formation of instabilities mediated by radiative losses. We note that the formation of similar spatial features was also observed in previous laser experiments with shocks in argon [\cite{Loupias2009apss}], pointing to a common cause. It is worth noting the work by [\cite{Visco2012PRL}] and [\cite{Reighard2006revsci}] driving shocks in argon at a pressure of 1 bar however, besides these works, we were unable to find imaging results that provide data on the shock morphology. 

The use of argon at a relatively high initial pressure (1 bar) and the resulting shock velocity of $\sim$40 km/s opens the question of the importance of radiative effects in these experiments. These effects can be quantified using the Boltzmann and Mihalas numbers following the definitions in [\cite{Falize2011ApJ}, \cite{Chaulagain2015HEDP}, \cite{Mabey2020}] and references therein. For the plasma conditions from our first simulations we estimate a Boltzmann number $\sim$1 and a Mihalas number of $\sim10^3$. The latter is unsurprising as only a handful of experiments in the past has claimed to be in a pressure dominated regime (see e.g. [\cite{Diziere2011apss}]). A Boltzmann number $\sim$1 implies that the radiative flow in our experiments is of the same order as the material flow which means we are in the threshold where the shock might not be strongly influenced by radiation. Looking at radiative effects from intermediate cases like this one to the case where the structure is dominated by radiation is necessary for a better understanding of the radiation of effect on shock waves as a general topic, which remains a very difficult topic at present.

Future work on SG-II will aim at providing more statistics of single shocks in argon and assess the role of radiative losses in the formation of these features. This will be complemented by a more detailed simulation work with the ARWEN code.

\begin{acknowledgments}

This research was supported by The Royal Society (UK) through a University Research Fellowship (URF-R-180032), a Research Fellows Enhancement Award (RGF-EA-180240), an International Exchanges grant (IES-R3-170140), and a Research Grant (RG2017-R2). The authors would like to thank the operation group of the SG-II laser facility. C.S.~acknowledges support from the French INSU-PNPS program. U.C.~acknowledges support by the project Advanced Research (CZ.02.1.01/0.0/0.0/16$\_$019/0000789) from European Regional Development Fund (ADONIS). F.S-V.~acknowledges the technical support from Paul Brown, Alan Finch and the entire team at the Physics Mechanical Instrumentation Workshop at Imperial College London.

\end{acknowledgments}

\bibliography{_2020_HPLSE_SG-II_60th_anniversary}

\end{document}